\documentclass[english,aps,pra,reprint,noshowpacs,superscriptaddress]{revtex4-1}   
\usepackage[T1]{fontenc}	
\usepackage[latin9]{inputenc}	
\usepackage{geometry}		
\usepackage{graphicx}
\usepackage[above,below]{placeins}	
\usepackage{times}
\usepackage{hyperref}  
\hypersetup{colorlinks=true,urlcolor=blue,citecolor=blue,linkcolor=blue}   
\newcommand\davidsays[1]{\textcolor{blue}{[\sc DR: {\em#1}]}}
\renewcommand\davidsays[1]{}

\begin{document}

\title{Paradigms in Physics 2.0}
\author{David Roundy}
\author{Elizabeth Gire}
\author{Ethan Minot}
\author{Emily van Zee}
\affiliation{Department of Physics, Oregon State University, Corvallis, Oregon, 97331}
\author{Corinne A. Manogue}
\affiliation{Department of Physics, Oregon State University, Corvallis, Oregon, 97331}


\begin{abstract}
In 2016, the Department of Physics at Oregon State University began a
process to revise our Paradigms in Physics curriculum for physics
majors.  We began with a colloquium to inform the department of our
plans and request their assistance, followed by a
survey of students and faculty as well as individual interviews with
the faculty teaching each course.  As we developed a plan to
address student- and faculty-identified challenges in the curriculum,
we met with each faculty member individually to explain and refine our
proposal, which was unanimously approved by the faculty.  Major
changes include major changes to several courses (math
methods, computational physics, modern physics, electronics, and
classical mechanics), including the introduction of two sophomore-year
courses designed specifically to help prepare students for their
upper-division courses.
\end{abstract}

\maketitle
%
The Paradigms in Physics project began in 1996, when three faculty
members at Oregon State applied to the NSF for funding to redesign the
upper-division curriculum.  The motivation for the original Paradigms~1.0
project was in fact similar to the motivation for Paradigms~2.0: a
desire to soften the ``brick wall'' encountered by students upon
reaching the junior year, within the constraint that transfer students
must be able to graduate with two years of upper-division
courses~\cite{manogue2001paradigms}.  On top of this, there was a
desire to provide students with a broad knowledge of physics prior to
the GRE exams in the fall of their senior year.  The process of
redesigning the curriculum was guided by the creation of index cards
listing subject content, and sorting these cards into courses.  This
process involved the entire faculty, and culminated in a unanimous
agreement to adopt the resulting curriculum.

The resulting curriculum was primarily composed of intensive junior-year \emph{Paradigm}
courses, followed by more conventional senior-year courses.
The \emph{Paradigm} courses meet every day for a total of seven
hours per week for 3 weeks.  These courses incorporate
laboratory experiences and active engagement into the class, and
typically have two problem sets per week.  These courses also had
integrated math content, and were followed by a \emph{Math Methods} course.
The senior-year courses are more traditional 3-credit courses which meet
three hours per week for an entire 10-week quarter.

Two decades have passed since the original Paradigms~1.0 effort.
During this time we have made a number of changes: e.g. a new
first \emph{Paradigm} course was introduced to soften the beginning of the
junior year, the order of courses was shuffled more than once, \emph{Math
Methods} was moved from the beginning of the senior year to the end of
the junior year, and a computational laboratory course was introduced
to accompany the junior-year \emph{Paradigm} courses.  The faculty maintained the tradition of meeting
every three weeks to discuss issues relating to upper-division
teaching.  New faculty arrived and learned to teach the new courses,
and introduced their own ideas to the courses.

A number of factors motivated us to embark on the
Paradigms~2.0 process.  In the last two decades, we have observed
a number of challenges students face in our major.  While in many cases we
addressed these by changing and reordering the courses, we felt a look
at the entire curriculum was in order.
In addition, our faculty
understanding of the details of the sequence was diminishing by attrition: 
while our younger
faculty are enthusiastic about the Paradigms program, many lack perspective on
where students are in a given course, and what content is essential
for students to grasp for a subsequent course.

\paragraph{Transfer students}
Our curriculum is strongly affected by our desire to accommodate
transfer students from community colleges.  This leads us to focus on
a physics major in which all upper-division courses are taken in two
years.  However, our current curriculum is very hard on students in
the fall of their junior year, and especially so for transfer
students.  So we aimed to structure and order the courses such that non-transfer
students could be better prepared for their junior year, while
transfer students could postpone a few courses so that the
Fall quarter of their junior year would be no more difficult than that of any other student.

\paragraph{Coupled curricular changes}
A backlog of curricular changes had accumulated that could not be
separated from an examination of the curriculum as a whole.  The requirements
for computational physics, the number of required electronics credits, and the
role of our modern physics course could not easily be addressed separately.

\paragraph{New faculty}

Only 5 of our 16 current tenure-line faculty were involved in the
original Paradigms~1.0 effort.  Some of the newest faculty have only a
superficial understanding of our course structure.  This poses a
challenge when these professors teach courses that are closely
intertwined with one another.

\section{Paradigms 2.0 process}
\begin{figure}
\includegraphics[width=\columnwidth]{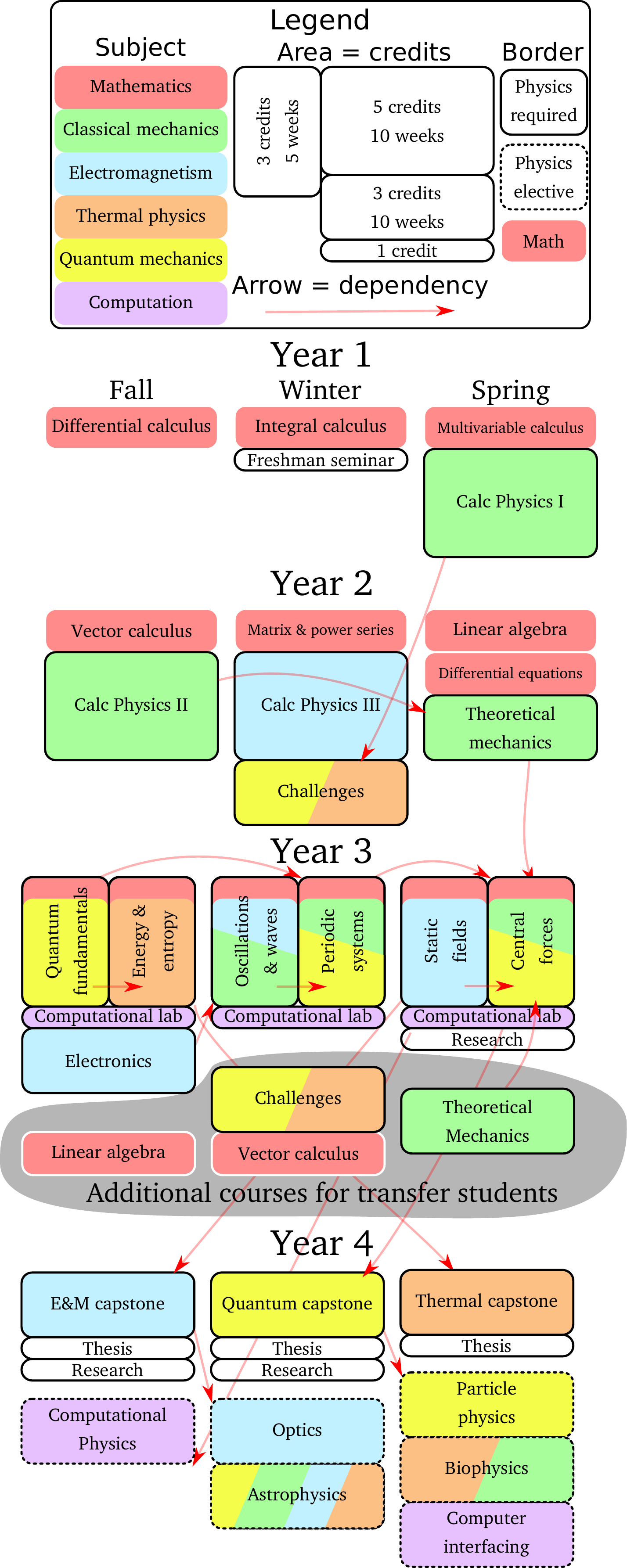}
\caption{Hypothetical student schedule in the new curriculum.
  Transfer students begin in year 3, and take the marked additional
  courses.\label{fig:schedule}}
\end{figure}

Like the process that led to Paradigms 1.0, the Paradigms 2.0 process is a ``shared vision'' type of change \cite{henderson2010}. Changes to the course structures of the physics curriculum emerged from discussions among the members of the department. We began the Paradigms~2.0 process in the Winter of 2016, having in
the previous year obtained faculty agreement, and support from our
Department head.  The process was spearheaded by a committee of four
(the authors DR, EG, EM, and CAM, with EvZ present at meetings
documenting our process).  During the Winter quarter, we informed the
community of our process through a colloquium, and collected student
and faculty perspectives on the existing curriculum.  We then
interviewed faculty who recently taught each of our existing courses
to document topics that were currently covered.  This resulted in a
total of $\sim$700 index cards in $\sim$30 stacks, with each stack
corresponding to one course, and each card describing a topic,
color-coded as in Fig.~\ref{fig:schedule}.

The committee met twice a week to discuss existing challenges and sort
the cards into new stacks corresponding to new and reordered courses.
When we discussed major changes to a course, we often invited
interested faculty to join us to provide their own perspective on
possible challenges and improvements.  Once we had a draft proposal,
we began inviting each faculty member in to see the cards and discuss
the proposed sequence of courses.  In addition, we had a focus group with
all the current students to explain the proposed changes and request
feedback.

After incorporating the feedback from  individual meetings with every
faculty member, we scheduled
two full faculty meetings with a week in between.  In the first meeting,
we presented our proposal in detail and invited questions, but not discussion.
This was needed for a couple of reasons:  those faculty we met with first
may not have seen the final proposal, and some faculty during their
individual meetings chose to focus on a small subset of the
curriculum, often relating to courses they had themselves taught.
During the following week faculty engaged in hallway discussions of the proposal.
This week helped ensure that the faculty did not feel rushed into a vote.
In the second faculty meeting, we again presented our proposal, and opened
the floor for discussion.  After considerable discussion, largely on
changes that were not part of the proposal, the faculty unanimously voted to adopt
the proposed changes.

\section{Changes made}
As a result of this process, we have implemented a number of changes
to our physics major curriculum.  The resulting curriculum is outlined
in Fig.~\ref{fig:schedule}.  The changes consisted of introducing two
new sophomore-level courses (which may be taken in the junior year) to
better prepare our majors for the their junior year.  We eliminated \emph{Modern
Physics} and the \emph{Classical Mechanics Capstone} in favor of the two new
sophomore-level courses.  We eliminated the \emph{Math Methods} course in
favor of \emph{Math Bits} integrated into the \emph{Paradigm} courses.
Finally, we restructured our nine 3-week \emph{Paradigm} courses into six
5-week \emph{Paradigm} courses.  We changed our computational physics requirement,
and reduced the number of electronics courses.

\subsection{5-week \emph{Paradigm} courses}
To begin with the most distinctive feature of our curriculum, we
wanted to maintain the existing intensive 7-hour-per-week schedule in
the junior year, which we have found effective.  This schedule allows students to focus
intensely on a single topic, it fosters the building of a cohort of students, and 
the daily schedule is helpful for active engagement.  We chose to
change from three 3-week courses per quarter to two 5-week courses
per quarter.  This gives students who fall behind a chance to catch up, and enables
us to give more feedback to students prior to the final exam.  Having 3-credit courses
simplifies the setting of teaching loads, and by reducing the number of courses, 
we allow the curriculum to be taught with fewer faculty, albeit at an increased
work load per professor.

We put considerable thought into the content and ordering of the new
courses.  In most cases, we think of each 5-week \emph{Paradigm} course as either
one formerly 4-week \emph{Paradigm} course (the one that consumed the extra week
each quarter), two 3-week courses compressed, or three 3-week courses combined
into two 5-week paradigms.  The major
scheduling change is to place \emph{Static Fields} (electrostatics and
magnetostatics) in the spring quarter.  
This gives students, especially transfer students, 
a bit more time to take \emph{Vector Calculus}.  It also puts all use of
curvilinear coordinates in the spring quarter, which should ease the
learning of \emph{Central Forces}.  The final major change was to move
special relativity from the junior year to the new sophomore-level
\emph{Theoretical Mechanics} course.

\subsection{Math bits}
Since Paradigms~1.0 was began, we introduced a
math-intensive week preceding two of the \emph{Paradigm} courses.  These math weeks provided
just-in-time preparation of the math skills required for those courses.
We have found these math weeks to be effective and popular with students.  In contrast to
these weeks the \emph{Math Methods} course is unwieldy, and challenging to place in
the curriculum at a time where it is helpful to students.  Moreover, some of the content
in \emph{Math Methods} was not actually required for our undergraduate courses, and only
needed for students bound for graduate school.  We chose therefore to eliminate the \emph{Math Methods} course in favor of \emph{Math Bits} consisting of a single week of 
just-in-time math content incorporated in each \emph{Paradigm} course.  Advanced
students may also take our graduate-level \emph{Math Methods} course.  
The \emph{Math Bits} for the entire year is taught by a single professor,
which provides continuity and coherence for the students.  
This provides some assistance for
professors teaching \emph{Paradigm} courses, and at the same time ensures
that they do not succumb to the temptation to short-change the math content
to the detriment of the students.

\subsection{Electronics}
We chose to reduce the \emph{Electronics} requirement from two 3-credit
courses to one 3-credit course.  For many physics majors, this is more
than sufficient, and gives students a greater number of electives.  In
addition, we removed the lecture section of this course---which had
developed an unusually high student work load for a 3-credit
course---in favor of in-lab instruction.

A final change to Electronics is that we now will \emph{require}
\emph{Electronics} during the junior year (specifically as a prerequisite for
\emph{Oscillations and Waves}).  This is made possible by the reduction in
Fall workload for incoming transfer students, who had
usually taken \emph{Electronics} as a senior.  This change has enabled us to
articulate distinct and sequenced learning outcomes from these two
courses, particularly in the realm of complex exponentials and Fourier
transforms.  It is also beneficial in providing students with
trouble-shooting skills prior to the in-class labs taught in
\emph{Oscillations and Waves}.

\subsection{Computational lab}
Over the last six years, we have been developing a 1-credit
computational laboratory course that accompanies the \emph{Paradigm} courses.  We
chose to require this course, while removing a requirement for a
lower-division 3-credit course in computational physics.  This
lower-division course was challenging to teach, since it always had a
mix of lower- and upper-division students, with very different skill
levels and needs.  Transfer students now take computation alongside
the non-transfer students. The course is taught in a laboratory setting using pair
programming~\cite{mcdowell2006pair} to help new programmers to learn.

\subsection{Sophomore courses}
We introduced two new sophomore-level courses: \emph{Physics of
Contemporary Challenges} and \emph{Techniques of Theoretical Mechanics}.  Both of
these courses ramp up student mathematical abilities prior to their
junior year.  The \emph{Challenges} course focuses on estimation,
dimensional reasoning, and interpretation of integrals, while
\emph{Theoretical Mechanics} teaches power series approximations and
exposes students to increased levels of mathematical sophistication and
sense-making strategies.

These courses have the challenge of teaching both juniors and sophomores
together.  They are taught in the winter and spring quarters, so
as to reduce the burden on transfer students in the Fall.  They were
explicitly constructed to \emph{not} teach any content required for
Fall or Winter junior-year courses so that they can be taken concurrently with
those courses.

\paragraph{Physics of Contemporary Challenges}
In this course we prepare students to
apply physics concepts and physical reasoning skills to sustainable
energy issues, climate change mechanisms, space exploration and
puzzles in fundamental physics. These ``real-world'' topics are chosen
for either the societal need (energy and climate), and/or the human
need to explore (space and fundamental physics). By prioritizing
inclusion of the most engaging challenges, we aim to attract and
retain as many potential physics majors as 
possible~\cite{kramer2016gathering}.

While contemporary challenges determine the narrative flow of the
course, physics concepts and physical reasoning skills are the main
substance. Students are introduced to thermodynamics, statistical
mechanics, electromagnetic radiation, quantum mechanics, and modern
experimental physics---in each case motivated by one or more
``contemporary challenges.'' Together with new physics concepts,
students are given new reasoning tools, such as the equipartition
theorem, the quantum-classical correspondence principle, order of
magnitude calculations, simplifying assumptions, and numerical
integration. Mathematically rigorous derivations are only briefly
mentioned in class. Instead we emphasize the physics concepts and reasoning
skills that allow professional physicists to quickly/quantitatively
make an initial assessment of a complex problem.

\paragraph{Techniques of Theoretical Mechanics}

While the \emph{Challenges} course takes a more experimental/applied physics
focus, \emph{Theoretical Mechanics} has a more
theoretical physics flavor. The theme of this course is the discussion
of strategies for making sense of physics problems and symbolic
problem solving. This sense-making includes coordinating and
interpreting symbolic expressions with conceptual understandings,
geometric relationships, and physical intuitions.

This course is aimed at students who are taking
or have completed the last course in the introductory sequence. The
physics content of the course is advanced Newtonian mechanics,
introduction to Lagrangian and Hamiltonian techniques, and special
relativity. These topics are convenient
for explicit discussion of sense-making strategies in physics because
(a) these students have recently studied problems that serve as limiting
cases for these more complex problems, (b) students at this level are
transitioning from solving problems primarily involving numbers to
problems with only symbolic parameters, and (c) sense-making as strategies 
can be discussed for approaching a problem, evaluating
answers, and refining intuitions about relativistic and other
unfamiliar situations. For example, we teach students how to use
spacetime diagrams to develop a story about a relativistic situation
and how to use hyperbola trigonometry on these diagrams to perform Lorentz
transformations that can be checked against the results of algebraic
Lorentz transformations.  The sense-making goals of the course are on
an equal footing with the physics goals, and sense-making is integrated
explicitly in all aspects of the course: exams, homework and during
in-class activities~\cite{lenz2017sensemaking, hahn2017sensemaking}. Our aim is that students will develop
sense-making skills that will improve their learning in advanced
courses and will be valued by their upper-division course instructors,
research advisors, and future employers.

\section{Summary}
We have developed a significant change to the Paradigms
curriculum.  This change focuses on the first two big ideas underlying
the Paradigms project: (1) close attention to content ordering, and (2)
developing a consensus and understanding of the curriculum among our
faculty members.  Our commitment to using active engagement in our
upper-division curriculum remains unchanged.  While the specific
content ordering we developed may be interesting, the \emph{process} by
which we reached it and developed a faculty consensus is at the
essence of what makes the Paradigms in Physics program distinctive.
One of the authors (EvZ) have completed a retrospective study of the 
process we used to develop the 
original Paradigms program and we have used her results extensively in 
designing and implementing the Paradigms~2.0 process.  She has continued 
to observe, study, and document our work as we have begun implementing 
the revised curriculum~\cite{lessonslearned}.
The process of developing faculty understanding of the Paradigms is
ongoing, as we are having professors new to the department shadow
experienced faculty as they teach \emph{Paradigm} courses, prior to teaching
the same course themselves.  

Future work will involve documenting the learning trajectories that we
have developed in our curriculum.  Furthermore, we are engaging in a
project to study our students' development of sense-making skills in
particular through the two new courses that are intended specifically
to ramp up those skills.

We look forward to another two decades of the Paradigms!


\acknowledgments{This work was supported by the National Science
  Foundation under Grant No. 1323800 Supplement.}

\bibliographystyle{apsrev}  	
\bibliography{paper}  	

\end{document}